\RequirePackage{fix-cm}
\RequirePackage{amsmath}

\documentclass[preprint, 12]{elsarticle}

\usepackage{graphicx}
\usepackage{fancybox}
\usepackage[T1]{fontenc}
\usepackage{empheq}
\usepackage{url}
\usepackage{wrapfig}
\usepackage{placeins}
\usepackage{subfig}
\usepackage[pscoord]{eso-pic}
\newcommand{\placetextbox}[3]{
  \setbox0=\hbox{#3}
  \AddToShipoutPictureFG*{
    \put(\LenToUnit{#1\paperwidth},\LenToUnit{#2\paperheight}){\vtop{{\null}\makebox[0pt][c]{#3}}}%
  }%
}%

\newcommand{\rpm}{\raisebox{.2ex}{$\scriptstyle\pm$}}

\newcommand{\specialcell}[2][c]{%
  \begin{tabular}[#1]{@{}c@{}}#2\end{tabular}}

\journal{Computer Networks}

\begin{document}
\begin{frontmatter}

\title{Multi-Gbps HTTP Traffic Analysis in Commodity Hardware Based on Local Knowledge of TCP Streams}
\placetextbox{0.5}{0.98}{\parbox{1.3\textwidth}{ \footnotesize{
    This is a PDF file of an unedited manuscript that has been accepted for publication. Please cite this article as: \\ \\
    \textbf{Carlos Vega, Paula Roquero, Javier Aracil, Multi-Gbps HTTP Traffic Analysis in Commodity Hardware Based on Local Knowledge of TCP Streams, Computer Networks (2017), doi: 10.1016/j.comnet.2017.01.001} \\ \\
    \textcopyright 2017. This manuscript version is made available under the CC-BY-NC-ND 4.0 license \\\url{http://creativecommons.org/licenses/by-nc-nd/4.0/}
}}}

\author[add1]{Carlos Vega}
\ead{carlosgonzalo.vega@predoc.uam.es}
\author[add1]{Paula Roquero}
\ead{paula.roquero@uam.es}
\author[add1]{Javier Aracil}
\ead{javier.aracil@uam.es}

\address[add1]{Departamento de Tecnología Electrónica y de las Comunicaciones, \\Escuela Politécnica Superior, Universidad Autónoma de Madrid. \\C/Francisco Tomás y Valiente 11 (28049) Madrid}

\date{Received: date / Accepted: date}


\begin{abstract}
In this paper we propose and implement novel techniques for performance evaluation of web traffic (response time, response code, etc.), with no reassembly of the underlying TCP connection, which severely restricts the traffic analysis throughput. Furthermore, our proposed software for HTTP traffic analysis runs in standard hardware, which is very cost-effective.
Besides, we present sub-TCP connection load balancing techniques that significantly increase throughput at the expense of losing very few HTTP transactions. Such techniques provide performance evaluation statistics which are indistinguishable from the single-threaded alternative with full TCP connection reassembly. 

\end{abstract}


\end{frontmatter}

\section{Introduction}
\label{sec:intro}

Large organizations such as banks, etc. make an increasing share of their business through the Internet \cite{ref:oecd}. Typically, HTTP is the protocol of choice to deliver services to the end-user, thanks to the widespread deployment of web clients in all kinds of mobile and desktop devices. Therefore, measuring the Quality of Service (QoS) provided by web portals \cite{ref:qoeqos} becomes of strategic importance. The same applies to other application protocols (VoIP, SIP, RTP, RTCP) \cite{ref:voip} but we focus on HTTP due to its widespread usage. Such QoS evaluation is normally based on response time statistics (from HTTP query to reply) and also on the analysis of response codes for the detection of anomalous behaviour in the monitored web services. For example, an HTTP error 500 indicates an internal server error, which must be taken care~of. 

The dissection and analysis of HTTP traffic can also be performed for cybersecurity purposes. However, the latter analysis is very fine-grain because security threats try to masquerade themselves among normal HTTP traffic. Therefore, losing a single HTTP transaction matters for security and forensic analysis. In contrast, the scope our research is network and service monitoring and not security, whereby only aggregated statistics such as means, averages or probability distributions matter.

Indeed, for QoS evaluation, only aggregate statistics are required, namely overall response time or percentage of a certain type of error codes. Furthermore, such statistics should be provided in real-time in order to timely react to possible anomalies. Once the overall statistics show performance degradation an in-depth analysis applies, which is normally performed off-line by inspecting the packets over a given time interval. In this light, the HTTP traffic analysis tool must be agile enough to cope with multi-Gb/s traffic rates and provide aggregate statistics in real-time, rather than providing a very high precision at the expense of a larger processing time. 

In this paper, we propose: 1) To lighten the underlying TCP connection reassembly and also to use a novel load balancing technique in order to sustain large offered traffic loads while keeping the accuracy at a reasonable level. 2)~With this, we provide real-time aggregate statistics of the processed HTTP traffic such as response time and response codes, among others. Furthermore, we have also attained a sustained 20 Gbps (2 x 10 Gbps) in a single host with several instances running in parallel.

The proposed techniques have been implemented in the {\em HTTPanalyzer} tool, as proof of concept and testbed for performance evaluation. Two real-world traces from large web commercial portals have been used to evaluate the maximum offered input traffic and the accuracy of the QoS statistics.  

The rest of the paper is organized as follows. First, prior to proceeding to the technical content of the paper we review the state of the art. Second, we describe the methodology and the proposed techniques for web traffic analysis, which are based on partial knowledge of the TCP connection, sub-TCP connection load-balancing and packet sampling. Finally, we discuss the performance evaluation and accuracy results, followed by the conclusions.

\begin{table}
    \centering
    \caption{Trace Files}
    \begin{tabular}{l|l|l|l}
    Capture File & Size   & Packets in file & HTTP Transactions \\ \hline
    trace1.pcap  & 387 GB & 539,178,347     & 13,743,811        \\ \hline
    trace2.pcap  & 120 GB & 211,823,223     & 3,681,812         \\
    \end{tabular}
     \label{tab:files}
\end{table}

\subsection{State of the art}
\label{sec:soa}
Most of the HTTP analysis tools available in the state of the art are more focused on reliability than on processing speed. Actually, some of them perform an offline analysis in which processing speed (henceforward throughput) is not a priority at all. Therefore, they are well-suited for cyber-security analysis of QoS evaluation offline, but not to the on-line analysis of a multi-Gbps stream to obtain real-time QoS metrics. Such tools are usually based on \textbf{TCP connection re-assembly} and, in a subsequent step, correlation of HTTP queries and responses in order to obtain the response time and error statistics. While this procedure provides very good results in terms of accuracy it adds a processing burden which makes it impossible to process data at very high speeds. 

We note that the HTTP 1.1 protocol is persistent, namely several HTTP requests and responses may be transmitted through the same TCP connection. Such requests and responses are then segmented in chunks and encapsulated within TCP segments. Thanks to the sequence numbers of each segment, the receiver can actually order the out-of-order segments and eliminate duplicates \cite{ref:tcp}. Then, TCP reassembles these segments into a data stream and feeds the application layer accordingly. Hence, reassembly of the TCP connection is a first mandatory step to retrieve each and every HTTP request and response at the receiver.

Due to this reassembly processing burden, such tools make use of sophisticated many-core and multicore systems to achieve high speeds. For example, Jing Xu et al. \cite{ref:tilera} propose a solution for dissecting HTTP traffic using the Tilera manycore platform for real-time HTTP traffic analysis up to 2 Gbps, which performs IP defragmentation and TCP reassembly. Even though the results are impressive we note that it requires a specialized PCI-e board for CPU offloading, in this case a Tilera TILEPro64 with 64 cores. We propose to use cost-effective ad-hoc hardware instead, at the expense of lesser accuracy, which still provides valuable statistics for the most network monitoring tasks either for online or offline analysis.

Another interesting example has been proposed by Kai Zhang et al. \cite{ref:http_parser} for a general purpose Intel multicore architecture, built on a pipelined RTC model, which also reassembles TCP connections, that achieves nearly 20 Gbps when parsing traffic looking for HTTP request and responses using 5 cores. According to their results, with a trace of 2,472,221 packets with an average length of 764 bytes the processing speed attains 3.3 Mpps. However, unlike our solution, requests and responses are not matched to obtain the desired HTTP performance statistics, for example response time.

Other tools like Bro, Fiddler, FlowScan, which do not run in specialised hardware, also provide a very high precision statistics at the expense of throughput, as they both reconstruct the whole TCP connection. Bro \cite{ref:bro} \cite{ref:bropaper} is a network security monitor that applies scripts and policies to events induced from packet streams, creating a series of logs and alerts about the network status. Fiddler \cite{ref:fiddler} is a HTTP debugging proxy server for Windows platforms that helps in the maintenance of web systems analysing traffic between the server and the browser. 

Furthermore, HTTPperf \cite{ref:httperf} is also a debugging tool that actively measures the performance of web services by generating HTTP traffic in order to test pre-production environments. FlowScan is a software package that analyzes NetFlow \cite{ref:netflow} data and provides graphical representations of different metrics. However, this tool may be overrun with the high number of flows of the analyzed traffic and \emph{``might not be able to scale beyond monitoring a couple fully-utilized OC3 (155 Mb/s) links.''} \cite{ref:flowscan} (p.314). Connection awareness requires a more complex processing and hence slower, since maintaining the status of thousands of different connections requires a large processing power. This is the general approach seen in different analysis tools from the state of the art.

For pure TCP reassembly tools, which can be used to extract the HTTP queries afterwards, {\em Libnids}, by Rafal Wojtczuk \cite{ref:libnids}, is a library, now discontinued, that provides TCP stream re-assembly and IP de-fragmentation, as well as TCP port scan detection in order to allow a deep analysis of TCP payloads like HTTP traffic among others. 

In conclusion, the state of the art shows that high-precision and throughput can only be obtained through specialized hardware. In this paper we provide a solution that trades-off high-precision and accuracy in ad-hoc hardware, which is inexpensive and easier to deploy and maintain, both for offline traces and online streams QoS analysis.

More specifically, the novelty of the paper is twofold. First, we propose a new HTTP analysis tool with a remarkable throughput by disregarding TCP flow reassembly. Second, we present a novel technique to distribute the HTTP traffic on a per transaction basis through multiple consumers, which increases throughput. Overall, our proposed techniques allow real-time analysis of high speed live HTTP traffic.

\section{Methodology}
\label{sec:methodology}

The traffic samples used for the experiments are described in Table \ref{tab:files} which consist of PCAP files made up of HTTP traffic from production proxies in two different large corporate networks with millions of HTTP transactions. We chose two different companies in order to have a larger and more diverse sample of this kind of traffic. Such files were used for assessing the accuracy and also for performance evaluation of our {\em HTTPanalyzer} tool.

As for accuracy evaluation, Tshark \cite{ref:tshark} \cite{ref:wireshark} was used as the ground truth reference, which is the de-facto traffic analysis tool nowadays. Such tool reassembles the TCP stack and uses multiple and complex packet dissectors for the different protocols available, providing detailed information of the traffic traces at the cost of slow processing speed. We note that Tshark is unable to process files of our traffic samples' size, due to its memory requirements, which are proportional to the file size. Consequently, we split up both samples in chunks of 20 million packets, which yields 27 chunks for trace1.pcap and 11 chunks for trace2.pcap with a size ranging from 13 to 15 GB. Since some transactions might be lost in the file boundaries, we also used the same files in chunks for our tool, for the sake of fairness.

\begin{figure}[!t]%
    \centering
    \subfloat[Scenario A: Offline test with a single consumer]{
      {\includegraphics[height=.26\columnwidth]{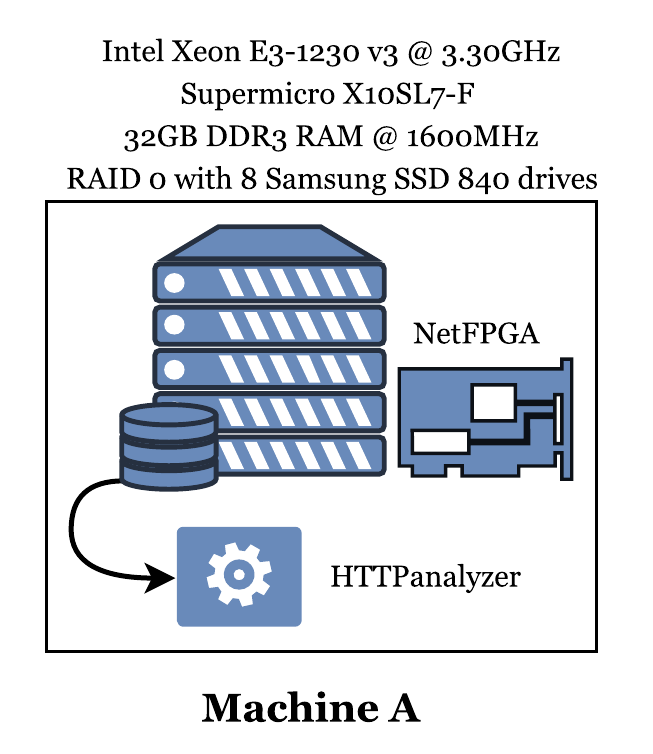} }
  \label{fig:scenarioa}
    }%
    \qquad
    \subfloat[Scenario B: Online test with multiple consumers]{
      {\includegraphics[height=.26\columnwidth]{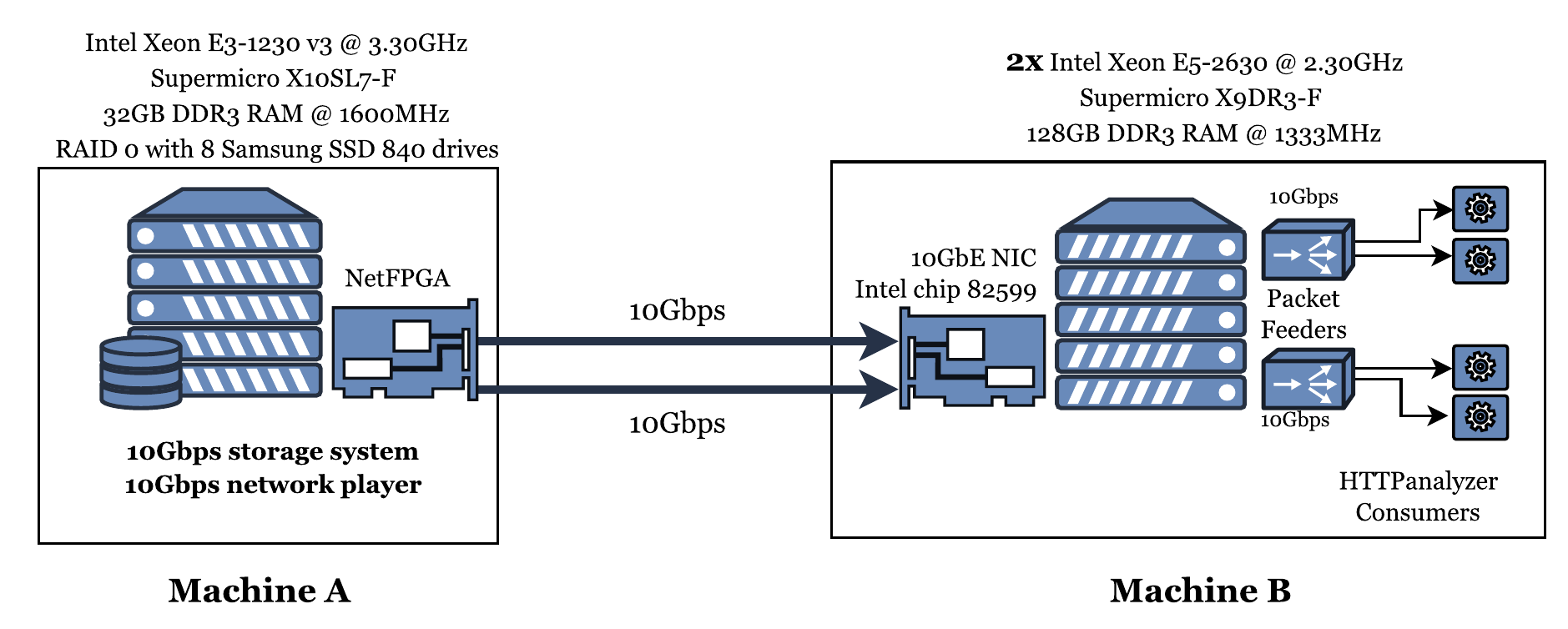} }
  \label{fig:scenariob}
  }%
    \caption{Different scenarios for both offline and online performance evaluation}%
    \label{fig:scenarios}%
\end{figure}

Regarding performance evaluation, we considered two different scenarios, for the assessment of accuracy and speed, respectively. The first one (offline, see Figure \ref{fig:scenarioa}) consisted of an offline processing of a trace file using a single instance of our tool, with the aim of comparing the accuracy of the results given by Tshark and our HTTPanalyzer. In the second scenario (online, see Figure \ref{fig:scenariob}), we employed several instances of HTTPanalyzer with a novel load balancing technique at the HTTP transactions level, instead of traditional TCP flow balancing, which is targeted for high speed processing at 20 Gbps. To this end, we used a balancer called packet feeder which receives the packets from the network interface and distributes them evenly to the HTTPanalyzer consumers through shared memory, while preserving the HTTP transaction consistency thanks to the hash functions that will be explained in section \ref{sec:loadbalance}. Namely, HTTP responses and their associated requests are sent to the same processing consumer. 

In section \ref{sec:results} we further discuss the trade-off between accuracy and speed  of HTTPanalyzer versus Tshark. 

\subsection{System modules}
The tool is structured internally in several modules, namely: a hash table for the HTTP requests and responses; two different pools of data structures for both the HTTP messages and the table cells, as well as a HTTP parsing module, among others. In the following sections we describe the proposed techniques for traffic dissection and analysis.

\subsubsection{Matching up HTTP requests and responses}
\label{sec:matching}

Offline traces are read using libpcap which supports packet filtering through Berkeley Packet Filter \cite{ref:bpf} with a default filter that passes through just the HTTP requests and responses. More specifically, the filter checks if the TCP payload begins with any HTTP method for the requests or the literal "HTTP" for the responses. This filter do not ensure the packet to be HTTP and it can be overridden by the user for its needs, hence, our tool checks the packet format during the packet dissection for further analysis.

Our {\em HTTPanalyzer} tool produces HTTP transaction records which are amenable to obtain valuable QoS statistics such as response time and response codes, among others. An example of a HTTP transaction record is showed next:
\begin{center}
\shadowbox{
\begin{minipage}{0.9\columnwidth}
123.111.50.23|2311|214.223.22.6|80|1393978285.777375000\\|1393978285.881505000|0.104130000|OK|200|\\GET|Mozilla/4.0|service.host.com|http://service.host.com/icon.gif
\end{minipage}
}
\end{center}
With the following format:
\begin{center}
\shadowbox{
\begin{minipage}{0.9\columnwidth}
client IP; client port; server IP; server port; request timestamp;\\response timestamp; response time; response message; response code; method; agent; host; URI
\end{minipage}
}
\end{center}
Interestingly, a key point of our dissection method is that \textbf{our tool does not reassembly the TCP connection}, and, furthermore, \textbf{only the first packet of the HTTP request} is considered for matching with the  corresponding HTTP response. Thus, only the first packet of the HTTP response is considered to obtain the HTTP transaction record. Therefore, we obtain the response time of the HTTP server as the time elapsed from the HTTP request to the HTTP response packets.

This way, we achieve notable speeds of 1.5 Mpps with a single instance of {\em HTTPanalyzer}. After the aforementioned filtering step, the HTTP request and response are extracted and passed to a correlation table. Thanks to a hash function, requests and responses are stored in their appropriate cell on a per transaction basis. Then, they are kept awaiting for their corresponding request or response to arrive and, should this happen, the transaction record is finally produced, in real-time.

\subsubsection{Hashing mechanism}
\label{sec:hash}

In what follows, we provide further insight into the hashing mechanism, which is the cornerstone for both high-speed processing and load balancing. The hash design is intrincate as it affects the hash table collision for the HTTP message processing as well as the load balancing of the traffic when using multiple consumers.

When storing HTTP requests and responses in the {\em HTTPanalyzer} tool, it becomes necessary to avoid collisions and to make an efficient use of the hash table. To do so, a uniform hash function is needed, also taking into account the speed restrictions we work with. Hash functions are also used to split the incoming packet stream evenly between consumers, and hence uniformity and randomness are key factors for the selection of a hash function.

As for \textbf{uniformity}, we aim to achieve the same probability for every hash value in the output range, thus reducing potential collisions. On the other hand, \textbf{randomness} \cite{ref:hashing} serves to distribute load between consumers, before the modulus operation is applied to determine where to send the packet. Actually, if packets are shared between two consumers, only the hash value least significant bit matters, i. e. whether it's equal to 0 or 1 with a probability close to 50\%. If not, the resulting packet streams will be unbalanced.

Generally speaking, only the 4-Tuple (see Equation 1) is used as a hash key to balance TCP flows, which ensures that packets from the same flow will end up in the same consumer. The problem with this approach is that some flows may carry more packets than others, leading to uneven packet distribution, and producing collisions when storing values in hash tables.

Even though the latter hash function provides uniformity, as it assigns an output value for each combination of input with the same probability, the real input values such as IP addresses and ports are not uniformly distributed on real datasets. For example, J.L. Garc{\'\i}a-Dorado et al. conclude \cite{ref:addrspatial} that they follow a Zipf distribution. Furthermore, as W. Shi et al. demonstrate \cite{ref:addrzipf}, owing to the Zipf-like distribution of the TCP flows, \textit{"a hash-based scheme is not able to achieve load balancing in parallel when the data follows a Zipf-like distribution"}.

\begin{equation}
\label{eq:4t}
Hash\; Value\; =\; Src.\; IP\; \oplus\; Src.\; Port\; \oplus\; Dst.\; IP\; \oplus\; Dst.\; Port
\end{equation}

\subsubsection{Reducing collisions on the hash table}
\label{sec:collisions}

In order to match HTTP requests and responses we do not need to ensure that all packets from the same flow end up in the same consumer. It is sufficient to ensure that at least both the request and its corresponding response reach the same consumer. Neither we need to store HTTP transactions on a per flow basis in our hash table, but rather per transaction.

Hence, our novel technique to circumvent this issue consist of a similar hash function but making use of  either the acknowledgement or sequence number. Such a hash function (see Equation \ref{eq:ack}) guarantees that HTTP messages from the same transaction will be stored on the same cell and will be distributed uniformly.

\begin{equation}
\label{eq:ack}
H.\; Value = \begin{cases}Request : & Src.\;IP \oplus Src.\;Port \oplus Dst.\;IP \oplus Dst.\;Port \oplus Ack\\Response : & Src.\;IP \oplus Src.\;Port \oplus Dst.\;IP \oplus Dst.\;Port \oplus Seq\end{cases}
\end{equation}

\subsubsection{Sub-TCP connection load balancing}
\label{sec:loadbalance}

For the parallel execution of multiple {\em HTTPanalyzer} consumers we use a load balancer tool (hereafter packet feeder) that distributes the packets between the {\em HTTPanalyzer} instances, reading the packets from the NIC's driver buffer and sharing a memory region with the consumers. For each incoming packet, a hash number is calculated using the packet headers and, then, the modulus operation is applied in order to choose the destination consumer for the packet. Using the generic 4-Tuple hash function (Equation \ref{eq:4t}) would ensure that packets from the same connection end up in the same consumer {\em HTTPanalyzer}. However, as noted before, such approach could lead to an unbalanced behaviour whenever some connections have a lot more packets and transactions than others.

Consequently, we use a similar function as Equation \ref{eq:ack}, but in order to achieve a better randomization of the least significant bits of the hash value, we also XOR up byte a byte this seq/ack number in addition to the previous operations. Then, we take the remainder of dividing this value by the number of consumers~(n), which yields the destination consumer. As a result, we obtain the hash function seen in Equation \ref{eq:balancer} which ensures that both consumers receive the same packet workload and that both requests and responses end up in the same consumer.

\begin{equation}
\label{eq:balancer}
Consumer = \begin{cases}Request: & Src. IP \oplus Src. Port \oplus Dst. IP \oplus Dst. Port \\ &\; \oplus\;Ack \oplus\;(Ack_1 \oplus Ack_2 \oplus Ack_3 \oplus Ack_4)
\\Response: & Src. IP \oplus Src. Port \oplus Dst. IP \oplus Dst. Port  \\ &\; \oplus\;Seq \oplus\; (Seq_1 \oplus Seq_2 \oplus Seq_3 \oplus Seq_4)\end{cases} mod.\; n
\end{equation}

In section \ref{sec:results} we discuss in detail the results of the proposed hash function and how well it distributes the hash values.

\subsubsection{Packet Processing}
\label{sec:pktprocessing}

\begin{figure}[!b]
\begin{center}
\includegraphics[width=\columnwidth]{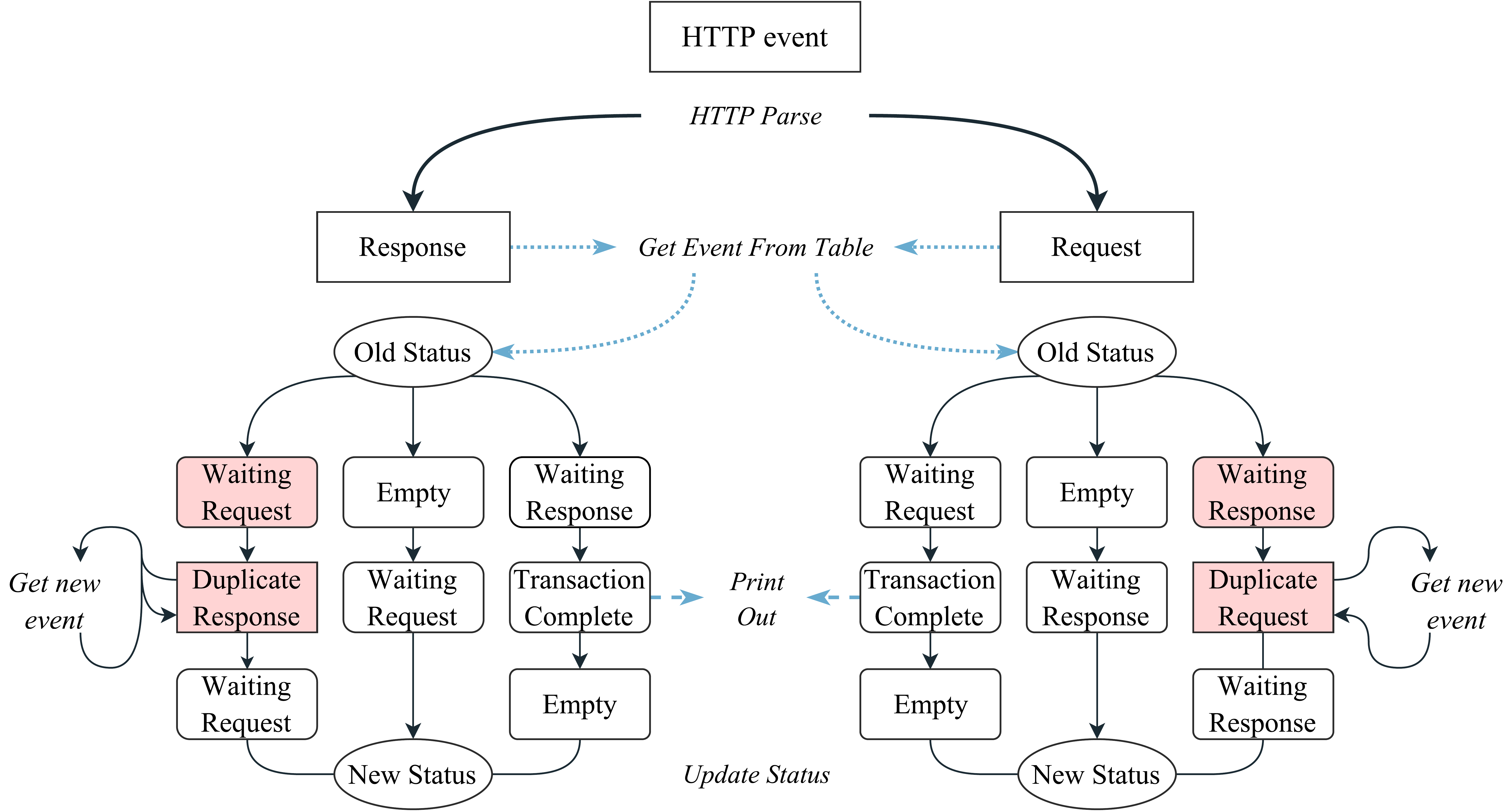}
\caption{Diagram of the processing of a HTTP message}
\label{fig:httpdiagram}
\end{center}
\end{figure}

As the Figure \ref{fig:httpdiagram} shows once the HTTP request or response arrives, a hash value is calculated by using the 4-Tuple formed by the source IP, source port, the destination IP and its corresponding destination port, as well as the acknowledgement number or sequence number depending on whether it is a request or response respectively (Equation 1). Such hash value is used to find the proper cell in the table by dividing it between the size of the table and taking the reminder. The main condition to pair an HTTP request with its response is that they both must match on their 4-Tuple (source IP, source port, destination IP, destination port) and \textbf{the HTTP response must have a sequence number equal to the HTTP request acknowledge number}. 

Afterwards, different possibilities arise depending on whether the cell may be either empty without a willing counterpart or taken by its suitor, which is awaiting. Nevertheless, there is a third scenario (showed in red in Figure \ref{fig:httpdiagram}) in which a duplicate message is already stored, being this message either a request or response that has arrived before. Mostly these cases mean candidate retransmissions or duplicates but an special case happens for the {\em 100 Continue} HTTP responses which usually happen during long POST HTTP requests. Such long requests normally end with a final response code (200 OK on a successful event) at the end of the transaction. We store such duplicates on the table as collisions looking forward to find its retransmitted/duplicated counterpart. Should the latter not arrive, they are cleaned from the hash table by the garbage collector.

\subsection{Limitations due to partial knowledge of the TCP connection}
\label{sec:limitations}

We also note that the aforementioned procedure is not as precise as the complete reassembly of the TCP flows due to packet misordering and retransmissions. Namely, we are using partial knowledge of the TCP connection at the vicinity of each HTTP transaction, and not global knowledge of the entire TCP connection. While this is advantageous for speed, there are indeed limitations for accurately extracting HTTP requests and responses from the TCP connection. However, we have used several heuristics to mitigate such inaccuracies as much as possible, which are presented next. 

\subsubsection{Unordered HTTP messages}
\begin{wrapfigure}[13]{r}[-2pt]{0.3\columnwidth}
\vspace{-1cm}
\centering
\includegraphics[width=0.3\columnwidth]{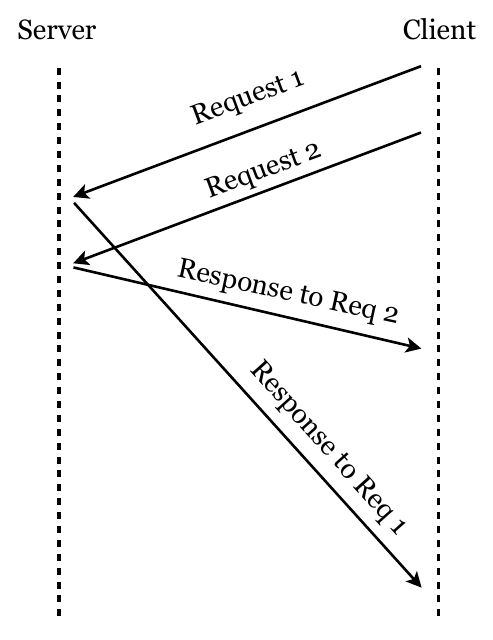}
\caption{Messages may arrive unordered}
\label{fig:unorderedTCP}
\end{wrapfigure}
First, the HTTP messages may arrive unordered, implying that a response corresponding to an older request can actually arrive later than a response to a more recent request (within the same TCP connection) as shown in figure \ref{fig:unorderedTCP}. Namely, HTTP transactions may be printed out of order. This is because the TCP connection is not reassembled, and thus, TCP segments may arrive in arbitrary order depending on the IP packet dynamics along the route from client to server. To partially circumvent this issue we do store the HTTP message whether it is a request or response and keep it waiting to the counterpart, hence, pairing can happen in both orders.

\subsubsection{Retransmissions}
Retransmissions are more frequent than unordered packets, resulting in duplicate transactions records. In the event of retransmitted messages, they are stored on their corresponding cell as well, in the collision list, resulting in duplicate transactions records. Such duplicate records   must be filtered out afterwards by the analyst, by looking for HTTP transactions with the same 4-tuple and ack/seq number.

\subsubsection{Accuracy}
We are well aware that full accuracy in detecting HTTP requests and responses is not possible with our approximate method. However, the aim of our research is to extract aggregate statistics that are amenable to use in a Network Operations Center (NOC), thus sacrificing accuracy for speed. 

For example, as explained before, only the first packet of the request and response is considered in the evaluation of response time and response codes. Thus, the URL might be truncated if the packet is longer than the MTU (1,518 bytes). The RFC 2616 (Hypertext Transfer Protocol HTTP/1.1) section~3.2.1~\cite{ref:rfchttp} says that \emph{``The HTTP protocol does not place any a priori limit on the length of a URI. Servers MUST be able to handle the URI of any resource they serve''} but the truth is that most browsers~\cite{ref:urllength} support 80,000 characters in average and the Apache Server has a limit of 8,192. 

Some browsers like Internet Explorer have a limit of 2,048 characters. Furthermore, large URLs are not good if web services intend to be indexed by search engines because the sitemaps protocol~\cite{ref:sitemaps} has a limit of 2,048 characters for the URL and SEO systems give less credit to these URLs.

In the results section we will show that the aggregate statistics obtained through our proposed technique are almost the same from those obtained with full TCP connection reassembly, and with a very high throughput.

\subsubsection{Garbage Collector}

Chances are that some of the requests and responses will never be removed from the hash table if the corresponding counterpart is not present in the trace, which entails wasting resources and possibly gives rise to collisions in the hash table. The same happens for very delayed responses, whose associated request occupies resources for too long. Both effects jeopardise throughput because the larger the hash table the larger the search time to find the appropriate cell.

To mitigate these effects, a garbage collector checks the state of the HTTP records' table and goes through all the active cells in the hash table removing transactions that shown no changes during the last 60 seconds of capture. Such unmatched HTTP messages are printed out together with the rest of HTTP transactions because they are valuable information for the HTTP analysis as well. 

\section{Performance Evaluation}
\label{sec:results}

In this section we present the results and compare them with other existing solutions. Our main requirement is throughput, while keeping a reasonable level of accuracy for the HTTP performance statistics. We discuss accuracy issues first, namely data loss in the requested URL due to fragmentation in several packets, response times, response codes and HTTP operations. Finally, we provide the throughput results. 

\subsection{Accuracy tests}
\label{sec:accuracy}

The next subsections discuss the accuracy of the tool for the different metrics of the HTTP traffic statistics.

\subsubsection{Potential loss of data in the request URL}

\begin{figure}[!t]
\begin{center}
\includegraphics[width=\columnwidth]{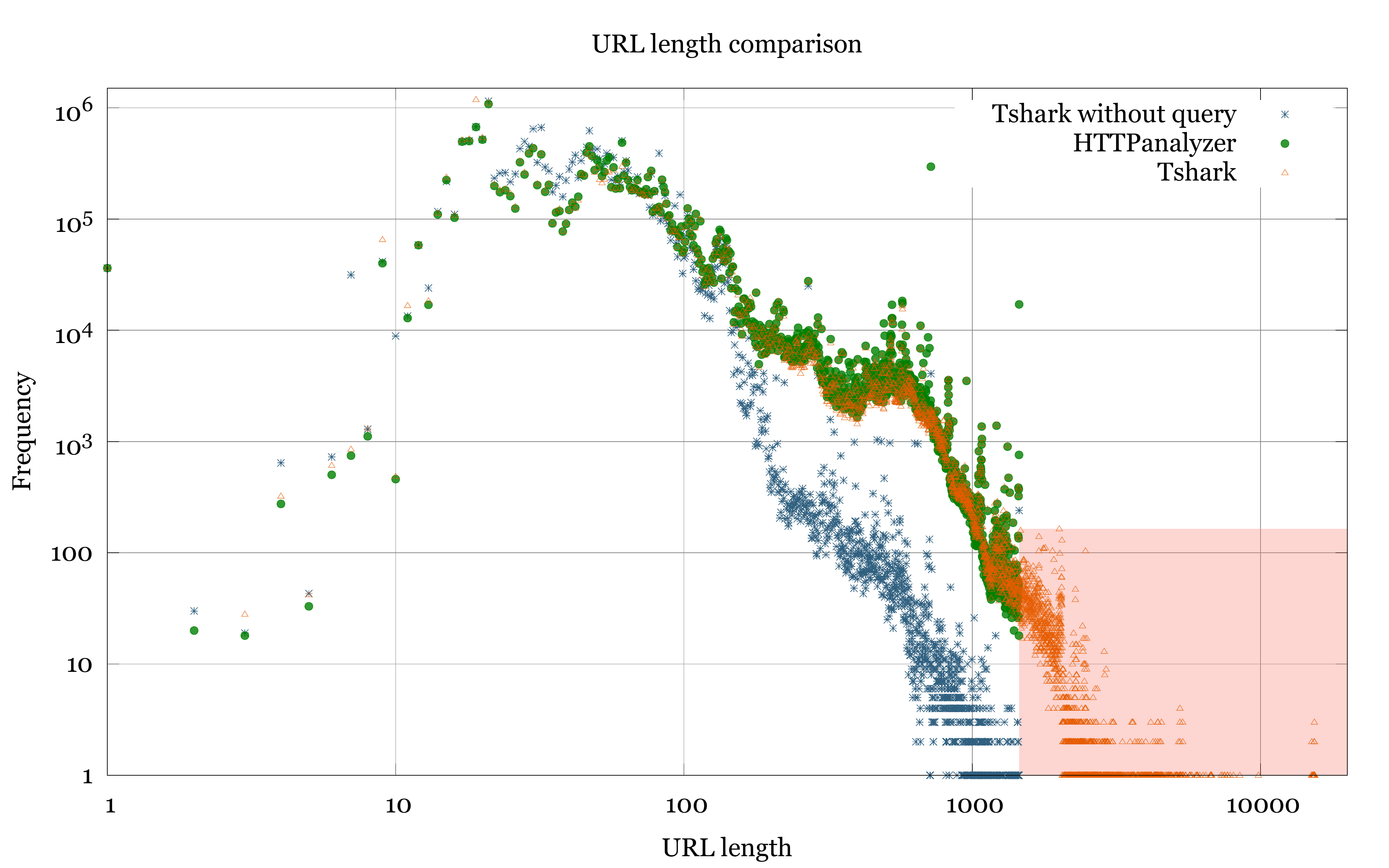}
\caption{URL size comparison. URLs in the area selected in red are longer than what our tool is able to manage, and represent a 0.04\% of the total URLs analyzed. }
\label{fig:url}
\end{center}
\end{figure}

For both our traffic samples, we studied (see Table \ref{tab:files}) how many URLs were truncated by our tool, and the maximum URL that was able to extract, and then compared it with the results given by Tshark. On Figure \ref{fig:url} we show that our tool (green circles) clearly matches Tshark results (showed in orange triangles), except for URLs over 1,455 characters, which is the maximum length our tool can manage. Such URLs are drawn in the chart as the points enclosed in the selected red area and represent only a 0.04\% of all URLs, considering both traces. 

Depending on the analysis performed, query parameters in the URL might be considered meaningful information or just query values that may be  discarded. We also drawn (in blue asterisks) the Tshark results disregarding URL query parameters and found that none of them exceeded our 1,455 character limit, showing that most of the URL length is composed of these query parameters. We believe that the most meaningful part of the URL is actually at the beginning, that shows the invoked resource, rather than the parameters afterwards. In any case, the HTTP transaction record contains enough parameters (4-tuple, time) to easily filter the packets corresponding to "long URLs" and, eventually, proceed to manual analysis.

\subsubsection{Response Time}

The response time is one of the most interesting HTTP QoS metrics, which serves to detect sudden degradation of Web services. We have compared the response time Complementary Cumulative Distribution Function (CCDF) using HTTP transaction response time data from  Tshark and our tool. 

Our tool measures this response time as the difference between the timestamp of the first packet of the HTTP request and the arrival time of the first packet of the response. However, Tshark usually measures HTTP response time as the time between the first request packet and the last packet of the response.

\begin{figure}[!t]
\begin{center}

\includegraphics[width=\columnwidth]{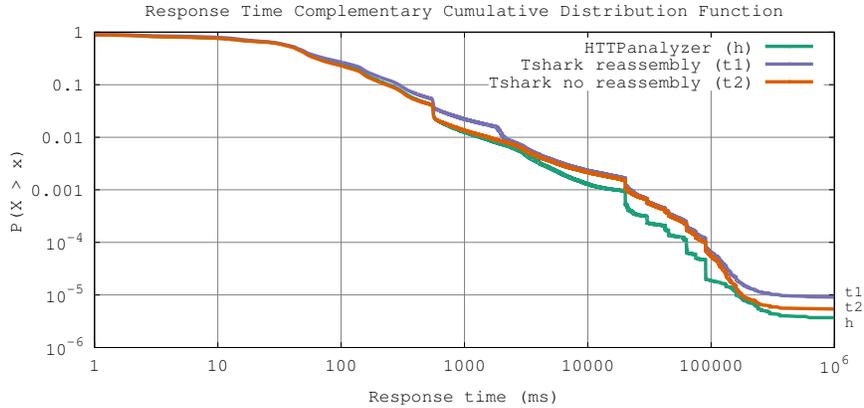}
\caption{Accuracy comparison between Response Time CCDFs}
\label{fig:ccdf}

\end{center}
\end{figure}

Notwithstanding, Tshark is also able to measure the response time in a different fashion when TCP reassembly is disabled using just the first packets as we do. 
Hence, in order to make a fair comparison, we present in Figure \ref{fig:ccdf} the results of both measure modes of Tshark together with our tool results, showing that there is no significant difference for this metric.

\subsubsection{Response codes}

\begin{figure}[!t]
\centering
\includegraphics[width=.85\columnwidth]{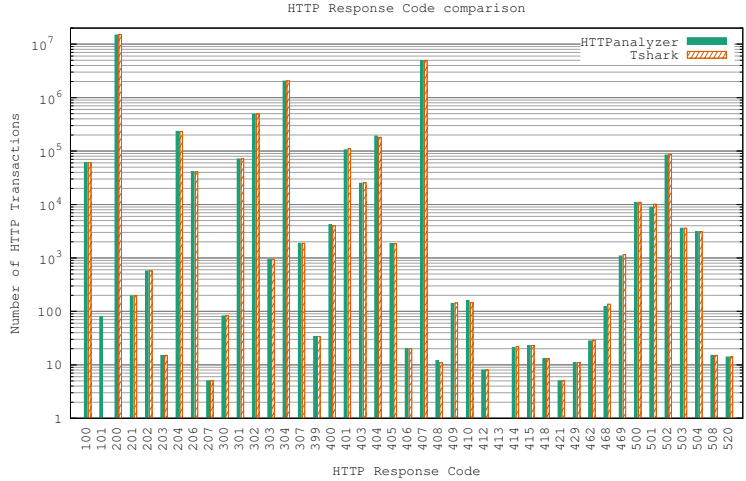}
\caption{HTTP Response Code Counting}
\label{fig:codes}
\end{figure}

The evaluation of the response codes is of fundamental importance to find errors. For example, a large number of 404 (Not Found) status codes implies that dead links may be found in the website or that an specific resource has disappeared. On the other hand, 5xx error codes are also of interest, such as the 500 code (Internal error), which may be delivered frequently by dynamic webs in case of failure in the dynamic objects invoked.

As Figure \ref{fig:codes} shows, the response code count is almost identical to the Tshark results, and the average count difference with Tshark for some specific response codes is 2.6\% with a median of 1.3\%. This difference is due to some loss in {\em HTTPanalyzer} when multiple requests are sent pipelined in the flow.

\subsubsection{HTTP Methods}

\begin{figure}[!b]
\centering
\includegraphics[width=0.75\columnwidth]{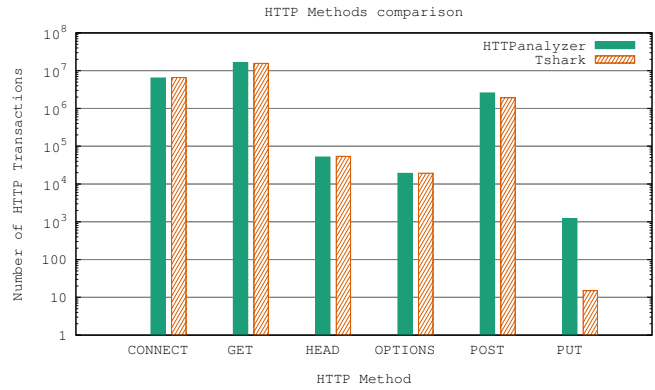}
\caption{HTTP Method Counting}
\label{fig:methods}
\end{figure}

A similar comparison can be done with the other HTTP transactions' statistics like the histogram of request methods in Figure \ref{fig:methods}, that shows that our tool provides nearly the same number of HTTP verbs as Tshark in the processed capture file. Some slight differences like the Tshark counting of the PUT method are due to the lost transactions in the file boundaries between chunks. As it turns out, we had to split our trace files into smaller chunks for Tshark to process them. Otherwise, the file size was too big and Tshark could not complete execution. 

\subsection{Throughput tests}

In this section, the throughput experiments have also been performed with both files from table \ref{tab:files}, in two different scenarios. First we have conducted offline tests in order to test the ability to process the sample files using high speed storage systems at 10Gbps with a single instance of the {\em HTTPanalyzer}. Then, we assessed the performance of our tool when processing 20 Gbps of live traffic sent with a traffic player from one host and receiving it on another, which in turn incorporates our {\em packetFeeder} software load balancer in order to split the incoming traffic between instances of our tool, making use of a uniform hash function.

However, in order to better understand the results, let us provide some more insight into the hash function used to distribute the packets both on the HTTPanalyzer hash table and between consumers.

\subsubsection{Hash function tests}

\begin{figure}[!b]
\centering
\includegraphics[width=0.8\columnwidth]{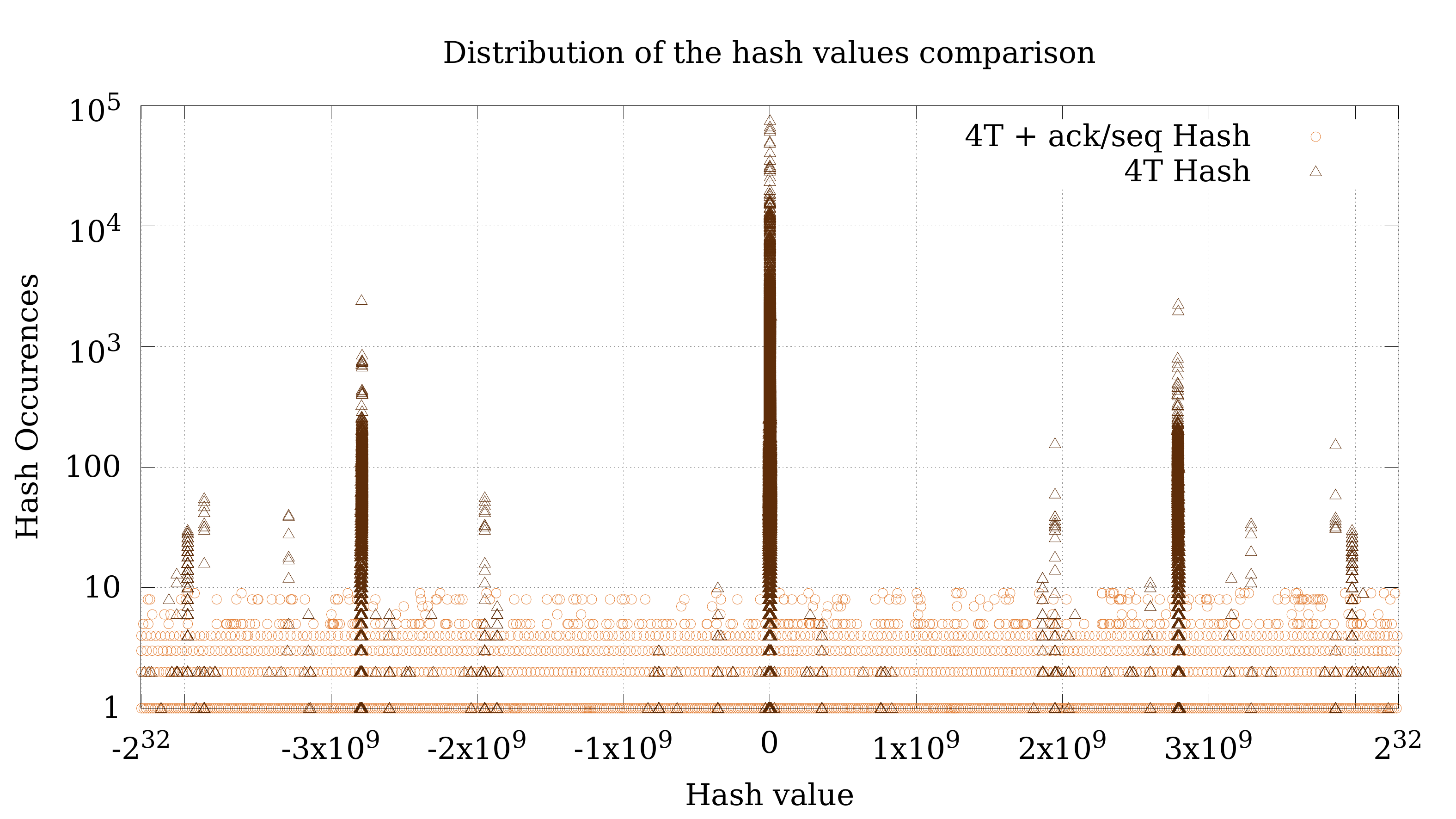}
\caption{Comparison of the distribution of the packets using different hash functions}
\label{fig:hash}
\end{figure}

The hash value histogram is the figure of merit for hash selection, as it summarizes, in a single graph, if the hash value is uniform for an even packet load balancing. In this light, Figure \ref{fig:hash} shows the hash value histogram for the Equations \ref{eq:4t} and \ref{eq:ack} explained in section \ref{sec:methodology}. We have divided this distribution in two sides with negative and positive values, meaning that the negative side of the X axis corresponds to one consumer and the positive part to the other consumer. Each point represents the number of occurrences for an specific hash value. 

Interestingly, we note that 32 bit numbers (sequence and ACK numbers length in TCP), adding up the sequence (seq) or acknowledge (ack) number accordingly, randomizes the resulting hash value reducing collisions and without affecting the pairing task. This refers to the previously explained hash function seen in section \ref{sec:methodology} on Equation \ref{eq:ack}.

As the Figure \ref{fig:hash} shows, collisions are largely reduced when using the seq and ack numbers because these 32 bit numbers randomize the entire hash, and they are initialized randomly by the TCP stack when flows are created. This distribution (shown in light brown with circles) is the same for both Equation \ref{eq:ack} and \ref{eq:balancer}. However, we used Equation \ref{eq:balancer} for the multi-consumer experiments while Equation \ref{eq:ack} will be used in the HTTPanalyzer hash table. We note that in the hash table there is no need for the least significant bits to be random and, consequently, we reduce the processing requirements to compute the hash.

In the light of the above discussion, we proceed with the presentation of the throughput results of the offline and online scenarios (see Figure \ref{fig:scenarios}).

\subsubsection{Single consumer tests}
\label{sec:single_tests}

\begin{table}[!t]
	\centering
    \caption{HTTPanalyzer speed benchmarks}
    \begin{tabular}{l|l|l}
    Storage System & Speed (Gbps) & Speed (Mpps) \\ \hline
    RAID 0         & 10.6 \rpm 0.58  & 1.8 \rpm 0.15   \\ \hline
    RAM            & 13.8 \rpm 1.4   & 2.1 \rpm 0.26   \\ \hline
    \end{tabular}
   	\label{tab:scenario_a}
\end{table}

This first test aims to prove that {\em HTTPanalyzer} is able to dissect PCAP files at 10Gbps using high speed storage. Figure \ref{fig:scenarioa} represents this scenario. For this test we used an Intel Xeon~E3-1230~v3 @ 3.30Ghz with 32GB of RAM and a storage system formed by a~RAID~0 with 8 Samsung~840 SSD drives with read speeds higher than 10~Gbps. 

Tests were performed using the sample traffic files described in Table \ref{tab:files}. We also conducted an in-memory benchmark using a 15GB chunk of one of the original files stored in a RAM filesystem in order to measure the maximum speed of our tool. These tests gave successful results, (see Table \ref{tab:scenario_a}), showing that a single instance of our tool is able to process more than 10Gbps of traffic.

\subsubsection{Multi-consumer experiments}
\label{sec:multi_tests}

This subsection discusses the results of the tests conducted using multiple HTTPanalyzer consumers for processing 20Gbps (two 10 Gbps streams) of online traffic on a single host. Our aim is to prove that many different instances of HTTPanalyzer can work in parallel with a similar load thanks to our hash implementation, with the benefit of achieving multi-Gbps throughput in a single host. To perform the experiment, two hosts were used as shown in Figure \ref{fig:scenariob}. Host A is the same server used for the previous scenario, but this time, the traffic samples stored on the high speed RAID system were sent using a NetFPGA traffic player \cite{ref:generador} across two 10GbE optic links, sending the same data through each cable. This 10G Trace Tester \cite{ref:tracetester} is a testbed part of the european project Fed4Fire able to send traffic at 10Gbps per link.

Right after, host B receives the traffic using HPCAP \cite{ref:hpcap}, a kernel-level driver designed for Intel NICs aiming to process a fully saturated 10GbE link. Since the driver reads the packets from each interface separately  two instances of the packet feeder were used, one for each 10GbE line; and for each of these packet feeder instances, two {\em HTTPanalyzer} consumers were set. This makes a total of four {\em HTTPanalyzer} instance running in parallel on four different cores. Each packet feeder shared out the packets using the aforementioned hash function, which ensures a uniform distribution of packets and HTTP transactions per consumer.

\begin{table}[!t]
	\centering
    \caption{HTTPanalyzer consumers distribution results}
    \begin{tabular}{l|l|l|l|l}
    ~                & \specialcell{Consumer\\A-1} & \specialcell{Consumer\\A-2} & \specialcell{Consumer\\B-1} & \specialcell{Consumer\\B-2} \\ \hline
    Received Packets & 49.86\%      & 50.02\%      & 49.94\%      & 50.02\%      \\ \hline
    HTTP Transactions     & 50.01\%      & 49.98\%      & 50.01\%      & 49.98\%      \\
    \end{tabular}
    \label{tab:scenarioB}
\end{table}

Interestingly, all the four instances received roughly the same load, as Table~\ref{tab:scenarioB} shows. The results indicate that our proposed hash technique is very effective in load balancing.

Tests with 40GbE links could not be performed as this technology is yet minority and expensive, also owing to the limit of the traffic player that prevents us from testing higher speeds. However, these results show promise that our tool can handle higher rates using this very same approach of load sharing between multiple {\em HTTPanalyzer} instances. 

\subsubsection{Throughput comparison against Tshark}

\begin{figure}[!b]%
    \centering
    \subfloat[Processing speed comparison between HTTPanalyzer and Tshark]{
    	{\includegraphics[width=.35\columnwidth]{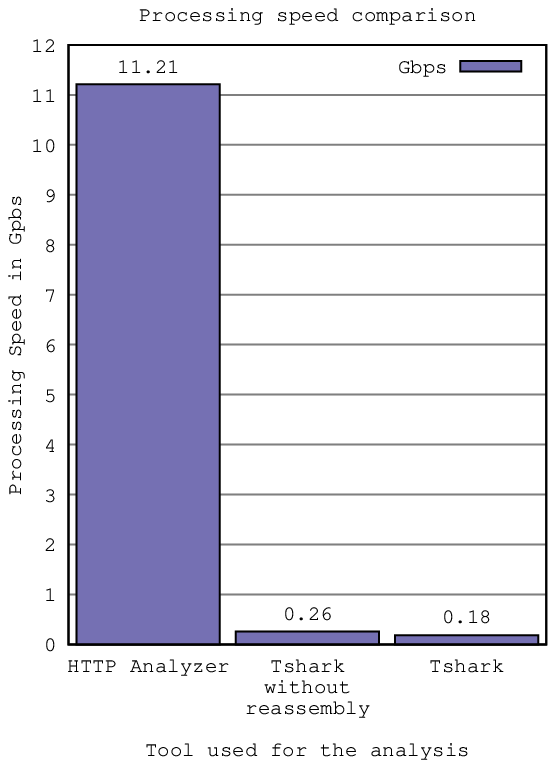} }
	\label{fig:tshark_speed}
    }%
    \qquad
    \subfloat[Percentage of packet loss of Tshark receiving online traffic from different traces]{
    	{\includegraphics[width=.35\columnwidth]{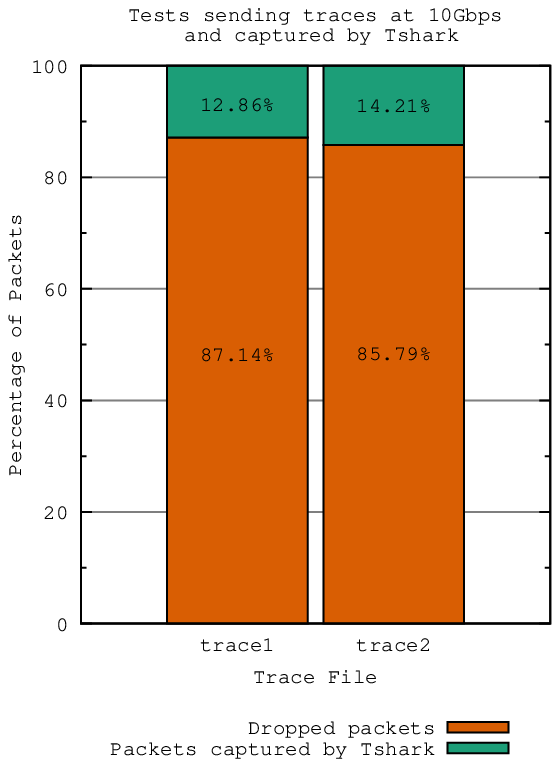} }
	\label{fig:tshark_loss}
	}%
    \caption{Performance charts of Tshark}%
    \label{fig:tshark_comparison}%
\end{figure}

To complete our throughput assessment, we compared the processing speed (or analysis throughput) of HTTPanalyzer versus Tshark. Even though Tshark provides highly detailed HTTP metrics, it turns out that it cannot cope with traffic sent at high speed for real-time analysis. 

Figure \ref{fig:tshark_speed} shows the processing speed of HTTPanalyzer and Tshark. It can be observed that HTTPanalyzer is 43 times faster than Tshark. The measurement experiment was performed offline, reading traces from a RAM filesystem as we did in \ref{sec:single_tests}. Furthermore, Figure \ref{fig:tshark_loss} represents the packet loss that Tshark suffered when traffic was injected at 10 Gbps speed during an online measurement experiment similar to those in section \ref{sec:multi_tests}. Clearly, the packet loss is very significant, which deems Tshark not adequate for on-line traffic analysis purposes in multi-Gbps scenarios. 
 
Actually, there is a trade-off between Tshark accuracy and HTTPanalyzer speed. However, the HTTPanalyzer accuracy is remarkable (as shown in \ref{sec:accuracy}), which, together with the throughput limitations of Tshark presented in this section, makes HTTPAnalyzer the tool of choice for real-time analysis of high speed traffic.

\section{Conclusions}

In this paper, we present a high-performance HTTP traffic analyzer that achieves 10 Gbps throughput with a single instance of the tool. A remarkable throughput of 20 Gbps online with live traffic can be achieved using multiple instances of the tool, thanks to our proposed hash function. All these results have been obtained in commodity hardware, with no need of ad-hoc high-speed network processors or massively parallel devices.

Finally, our tool provides real-time statistics of different aggregate metrics to measure the QoS of web traffic in large organizations. Such metrics are of strategic importance because of its close relation with the Quality of Experience of the final user, allowing to detect changes in the web services behaviour on-the-fly.

\section*{Acknowledgements}
The authors used the testbed 10G Trace Tester \cite{ref:tracetester}, which is part of the european project Fed4Fire under the aegis of the European Union's Seventh Framework Programme (FP7) under Grant FP7-ICT-318389.


\begin{thebibliography}{}

\bibitem{ref:oecd}
Pe{\~n}a-L{\'o}pez, Ismael and others: \emph{OECD Internet Economy Outlook 2012}, Chapter 4 (2012) \url{http://dx.doi.org/10.1787/9789264086463-en}

\bibitem{ref:qoeqos}
Stas Khirman and Peter Henriksen: \emph{Relationship between Quality-of-Service and Quality-of-Experience for Public Internet Service}. In In Proc. of the 3rd Workshop on Passive and Active Measurement. (2002) \\ 
\url{http://www-v1.icir.org/2002/Relationship_Between_QoS_and_QoE.pdf}

\bibitem{ref:voip}
Garc{\'\i}a-Dorado, Jos{\'e} Luis and Santiago del R{\'\i}o, Pedro M and Ramos, Javier and Muelas, David and Moreno, Victor and L{\'o}pez de Vergara, Jorge E and Aracil, Javier: \emph{Low-cost and high-performance: VoIP monitoring and full-data retention at multi-Gb/s rates using commodity hardware}. International Journal of Network Management. Vol. 24. (no. 3) 181-199. (2014) \url{http://dx.doi.org/10.1002/nem.1858}

\bibitem{ref:tcp}
Postel, J. (2003). RFC 793: Transmission control protocol, September 1981. Status: Standard, 88. \url{https://tools.ietf.org/html/rfc793#section-1.5} (Last accessed: 3 Dec. 2016)

\bibitem{ref:tilera}
Jing Xu; Hanbo Wang; Wei Liu; Xiaojun Hei: \emph{Towards High-Speed Real-Time HTTP Traffic Analysis on the Tilera Many-Core Platform}. IEEE HPCC\_EUC, (2013) \\
\url{http://dx.doi.org/10.1109/HPCC.and.EUC.2013.252}

\bibitem{ref:http_parser}
Kai Zhang; Junchang Wang; Bei Hua; Xinan Tang: \emph{Building High-performance Application Protocol Parsers on Multi-core Architectures}. IEEE 17th ICPADS. (2011) \\
\url{http://dx.doi.org/10.1109/ICPADS.2011.37}

\bibitem{ref:bro}
Bro.org: \emph{The Bro Network Security Monitor.} (2013) \url{http://www.bro.org} (Last accessed: 3 Dec. 2016)

\bibitem{ref:bropaper}
Vern Paxson, \emph{Bro: A System for Detecting Network Intruders in Real-Time.} International Journal of Computer and Telecommunications Networking (1998) \\ \url{https://doi.org/10.1016/S1389-1286(99)00112-7}

\bibitem{ref:fiddler}
Eric Lawrence: \emph{Debugging with Fiddler.} (2012) \url{https://fiddlerbook.com/book/} (Last accessed: 3 Dec. 2016)

\bibitem{ref:httperf}
Mosberger, D., and Jin, T. (1998). httperf ? a tool for measuring web server performance. ACM SIGMETRICS Performance Evaluation Review, 26(3), 31-37. \url{https://doi.org/10.1145/306225.306235}

\bibitem{ref:netflow}
NetFlow, Cisco IOS, White Paper, \emph{Introduction to Cisco IOS NetFlow-A Technical Overview} (2006) \url{http://www.cisco.com/c/en/us/products/collateral/ios-nx-os-software/ios-netflow/prod_white_paper0900aecd80406232.html}

\bibitem{ref:flowscan}
Plonka, David, in LISA, pp. 305-317, \emph{FlowScan: A Network Traffic Flow Reporting and Visualization Tool} (2000) \url{https://www.usenix.org/legacy/events/lisa2000/full_papers/plonka/plonka.pdf}

\bibitem{ref:libnids}
Rafal Wojtczuk: \emph{Libnids, an implementation of an E-component of Network Intrusion Detection System.} (2010) \url{http://libnids.sourceforge.net/} (Last accessed: 3 Dec. 2016)

\bibitem{ref:tshark}
Orebaugh, A., Ramirez, G., and Beale, J. (2006). Wireshark and Ethereal network protocol analyzer toolkit. Syngress.

\bibitem{ref:wireshark}
Combs, G. (2007). Wireshark. \url{http://www.wireshark.org/} (Last accessed: 3 Dec. 2016)

\bibitem{ref:bpf}
McCanne, S., and Jacobson, V. The BSD Packet Filter: A New Architecture for User-level Packet Capture. In USENIX winter (Vol. 46). (1993) \url{http://dl.acm.org/citation.cfm?id=1267305}

\bibitem{ref:hashing}
Korth, H. F., and Silberschatz, A : \emph{Database system concepts 6th Edition}. Chapter 11, page 510. (2010)

\bibitem{ref:addrspatial}
Garc{\'\i}a-Dorado, J. L., Hern{\'a}ndez, J. A., Aracil, J., de Vergara, J. E. L., Monserrat, F. J., Robles, E., and de Miguel, T. P: \emph{On the duration and spatial characteristics of Internet traffic measurement experiments}. IEEE Communications Magazine, Vol.46 issue 11, 148-155. (2008) \url{http://dx.doi.org/10.1109/MCOM.2008.4689258}

\bibitem{ref:addrzipf}
Shi, W., MacGregor, M. H., and Gburzynski, P. \emph{An adaptive load balancer for multiprocessor routers}. Simulation, Vol. 82 (no. 3), 173-192. (2006) \url{http://dx.doi.org/10.1177/0037549706067079}

\bibitem{ref:rfchttp}
Fielding, R., Gettys, J., Mogul, J., Frystyk, H., Masinter, L., Leach, P., and Berners-Lee, T. \emph{RFC 2616 Hypertext Transfer Protocol - HTTP/1.1 Hypertext Transfer Protocol.} (1999) , 1999. \url{http://www.w3.org/Protocols/rfc2616/rfc2616-sec3.html#sec3.2.1} 

\bibitem{ref:urllength}
\emph{Boutell.com: What is the maximum length of a URL?}. (2006) \\ \url{http://www.boutell.com/newfaq/misc/urllength.html} (Last accessed: 3 Dec. 2016)

\bibitem{ref:sitemaps}
Sitemaps.org: \emph{Sitemaps XML format.} (2008) \url{http://www.sitemaps.org/protocol.html} (Last accessed: 3 Dec. 2016)

\bibitem{ref:generador}
Jose Fernando Zazo, Marco Forconesi, Sergio Lopez-Buedo, Gustavo Sutter, and Javier Aracil: \emph{TNT10G: A high-accuracy 10 GbE traffic player and recorder for multi-Terabyte traces}. ReConFig14. (2014) \url{http://dx.doi.org/10.1109/ReConFig.2014.7032561}

\bibitem{ref:tracetester}
Jos{\'e} Luis Garc{\'\i}a-Dorado and Jose Fernando Zazo, HPCN-UAM: \emph{10Gbps Trace Reproduction testbed for testing software-defined networks (10GTRACE-TESTER)}. (2015) \\ \url{http://www.fed4fire.eu/10g-trace-tester/} (Last accessed: 3 Dec. 2016)

\bibitem{ref:hpcap}
Victor Moreno, Pedro M. Santiago del R{\'\i}o, Javier Ramos, David Muelas,  Jos{\'e} Luis Garc{\'\i}a-Dorado, Francisco J Gomez-Arribas, and Javier Aracil: \emph{Multi-granular, multi-purpose and multi-Gb/s monitoring on off-the-shelf systems}. International Journal of Network Management (2014) \\ \url{http://dx.doi.org/10.1002/nem.1861}

\end{thebibliography}


\end{document}